\documentclass[a4paper,11pt]{article}
\pdfoutput=1

\usepackage{amssymb,mathrsfs,amsmath}
\usepackage{a4wide}
\usepackage{color,xcolor}
\usepackage{slashed,soul}
\usepackage{graphicx}
\usepackage{caption}
\usepackage{subcaption}
\usepackage{amsfonts}
\usepackage{comment}
\usepackage{lscape}
\def\linkcolor{cyan!70!black}

\usepackage[
colorlinks=true
,urlcolor=\linkcolor
,anchorcolor=\linkcolor
,citecolor=\linkcolor
,filecolor=\linkcolor
,linkcolor=\linkcolor
,menucolor=\linkcolor
,linktocpage=true
,pdfproducer=medialab
,pdfa=true
]{hyperref}
\usepackage{amsthm}
\usepackage{booktabs}
\usepackage{array}
\usepackage{rotating}
\usepackage[numbers, sort&compress]{natbib}
\usepackage{multirow}
\usepackage{float}
\usepackage[utf8]{inputenc}
\usepackage[T1]{fontenc}
\usepackage{appendix}
\usepackage{orcidlink}
\usepackage{ragged2e}

\renewcommand{\baselinestretch}{1.2}

\let\OLDthebibliography\thebibliography
\renewcommand\thebibliography[1]{
  \OLDthebibliography{#1}
  \setlength{\parskip}{0pt}
  \setlength{\itemsep}{0pt plus 0.3ex}
}

\providecommand{\keywords}[1]
{
  \small	
  \textbf{\textit{Keywords:}} #1
}

\allowdisplaybreaks

\begin{document}

\vspace{1cm}

\vspace{0.8truecm}

\begin{center}
\renewcommand{\baselinestretch}{1.8}\normalsize
\boldmath
{\LARGE\textbf{Searching for the LFV $\gamma\gamma e\mu$ interaction at future $e^- e^+$ colliders
}}
\unboldmath
\end{center}

\vspace{0.4truecm}

\renewcommand*{\thefootnote}{\fnsymbol{footnote}}

\begin{center}
{\bf 
 M. A. Arroyo-Ure\~na$\,^{1\,,2}$\footnote{\href{mailto:marco.arroyo@fcfm.buap.mx}{marco.arroyo@fcfm.buap.mx}},
 R. Gait\'an$\,^3$\footnote{\href{mailto:rgaitan@unam.mx}{rgaitan@unam.mx}}, 
 Marcela Mar\'in$\,^3$\footnote{\href{mailto:marcemarino8a@cuautitlan.unam.mx}{marcemarino8a@cuautitlan.unam.mx}},
 Humberto Salazar$\,^{1\,,2}$\footnote{\href{mailto:hsalazar@fcfm.buap.mx}{hsalazar@fcfm.buap.mx}},
 M. G. Villanueva-Utrilla$\,^{1\,,2}$\footnote{\href{mailto:maria.villanueva@alumno.buap.mx}{maria.villanueva@alumno.buap.mx}},
\vspace{0.5truecm}}

{\footnotesize
$^1$ {\sl Facultad de Ciencias Físico-Matemáticas, Benemérita Universidad Autónoma de Puebla,  C.P. 72570, Puebla, M\'exico.\vspace{0.15truecm}}

$^2$ {\sl Centro Interdisciplinario de Investigación y Enseñanza de la Ciencia (CIIEC), Benemérita Universidad Autónoma de Puebla, C.P. 72570, Puebla, M\'exico.\vspace{0.15truecm}}

$^3$ {\sl Departamento de F\'isica, FES-Cuautitl\'an, UNAM, C.P. 54770, Estado de M\'exico, M\'exico.\vspace{0.15truecm}}

}
\end{center}

\renewcommand*{\thefootnote}{\arabic{footnote}}
\setcounter{footnote}{0}

\vspace{0.4cm}
\begin{abstract}
\noindent 

We investigate the Lepton Flavor Violating (LFV) process $e^{+}e^{-}\to e^{+}e^{-}e\mu$ ($e=e^-,\,e^+,\,\mu=\mu^-,\,\mu^+ $) at future circular colliders, probing the effective $\gamma\gamma e\mu$ interaction through photon fusion. Within an Effective Field Theory (EFT) framework compatible with theoretical and experimental constraints, we identify regions in the parameter space of the effective couplings where the signal could be observed at the Circular Electron-Positron Collider (CEPC) at $\sqrt{s}=240\ \text{GeV}$ and the Future Circular Collider (FCC-ee) at $\sqrt{s}=240$ and $350\ \text{GeV}$. Our analysis leverages distinctive kinematic distributions---particularly the transverse momentum of the scattered electron $p_{T}^{(e^{-}\ \text{scattered})}$ and the central muon $p_{T}^{\mu}$---to achieve efficient signal--background separation. By employing a neural network classifier, we enhance the sensitivity beyond traditional cut-based methods, demonstrating the discovery potential of these facilities for LFV searches in the clean environment of $e^{+}e^{-}$ collisions.

\end{abstract}


\section{Introduction}

In the Standard Model (SM), lepton flavor is strictly conserved in its original formulation with massless neutrinos. However, the observation of neutrino oscillations has firmly established that these particles possess nonzero masses~\cite{Super-Kamiokande:1998kpq,SNO:2001kpb,SNO:2002tuh}, implying lepton flavor violation in the neutral sector. Although the SM ---augmented by nonzero neutrino masses--- predicts charged LFV (cLFV) processes to occur at extremely suppressed rates due to the GIM mechanism~\cite{Petcov:1976ff}, any definitive experimental observation of cLFV would constitute incontrovertible evidence for physics beyond the SM. Over the past eight decades, extensive experimental efforts have been dedicated to searching for cLFV processes~\cite{Diociaiuti:2024stz,Blondel:2013ia,Belle-II:2018jsg} without a conclusive response. In this work, we explore the \( \gamma\gamma e\mu \) interaction, arising within a low-energy Effective Field Theory (EFT) framework, and assess its prospects for experimental investigation at the CEPC~\cite{Ai:2024nmn} and FCC-ee~\cite{FCC:2025lpp}.

Photon--photon interactions provide a remarkably clean probe of electromagnetic phenomena, as they are precisely described by Quantum Electrodynamics (QED) and efficiently modeled via the equivalent photon approximation (EPA)~\cite{Budnev:1975poe,vonWeizsacker:1934nji,Williams:1934ad}. At electron--positron colliders, the \(e^+\) and \(e^-\) beams serve as sources of quasi--real photons, enabling precise studies of processes such as Breit--Wheeler \(e^+e^-\) pair production~\cite{dEnterria:2013zqi,ATLAS:2019azn,CMS:2018erd} and exclusive dimuon production~\cite{CMS:2011vma,ATLAS:2015wnx}. These measurements provide stringent tests of QED and establish essential benchmarks for searches for rare phenomena, including charged lepton flavor violation in processes like \( \gamma\gamma \to e\mu\). Moreover, the exceptionally clean environment of electron--positron colliders —characterized by high luminosity and low background— significantly enhances sensitivity to the \( \gamma\gamma e\mu\) interaction.

In this context, CEPC is designed to reach up to $10.8$ ab$^{-1}$ at a center‑of‑mass energy of $\sqrt{s}= 240$~GeV~\cite{Ai:2024nmn}, while FCC-ee aims for up to $20.6$ ab$^{-1}$ at $\sqrt{s}=365$ GeV~\cite{FCC:2025lpp}. By contrast, the future linear colliders such as ILC~\cite{ ILCInternationalDevelopmentTeam:2022izu} and CLIC~\cite{Adli:2025swq}, despite the high center-of-mass energy they will achieve —$\sqrt{s}=1$ and $\sqrt{s}=3$ TeV, respectively— lack the integrated luminosity required to produce a sufficient number of signal events for probing the \(\gamma\gamma e\mu\) interaction with comparable sensitivity to the circular colliders. As will be shown later, high luminosities are more advantageous for the analysis than high center-of-mass energies. 

The theoretical framework adopted in this work can induce the radiative decays \(\ell_i \to \ell_j \gamma (\gamma)\), cLFV conversions in nuclei, three-body decays \(\ell_i \to \ell_j \bar{\ell}_k \ell_k\), and semileptonic tau decays \(\tau \to \ell PP\), where \(PP\) denotes a pair of light pseudoscalar mesons. A comparative analysis with current experimental bounds, as presented in Ref.~\cite{Fortuna:2023paj}, reveals that the most stringent constraints on the diphoton operators arise from the loop-induced radiative decays \( \ell_i \to \ell_j \gamma \)~\cite{MEG:2016leq,BaBar:2009hkt,Belle:2021ysv}. Nonetheless, despite the suppression of the relevant parameters, we identify a region of the parameter space in which the couplings that directly affect the proposed signal are favorable to motivate experimental searches for the cLFV process via photon--photon scattering at the aforementioned colliders.

This letter is structured as follows. In Sec.~\ref{Theory}, we present the theoretical framework in which our predictions are based. Experimental constraints on the model parameter space are also included. Section~\ref{Collider} focuses
on taking advantage of the insights gained from the previous section, performing a computational analysis of the proposed signal and its SM background processes. Finally, the conclusions are presented in Sec.~\ref{Conclusions}.

\section{Theoretical framework}\label{Theory}

The local charged lepton flavor violating interaction \( \gamma\gamma \bar{\ell}_i \ell_j \) is described by the effective Lagrangian~\cite{Bowman:1978kz, Davidson:2020ord}:
\begin{align}\label{Lagrangian_Q}
	\mathcal{L}_{\rm eff} &= \left( G_{SR}^{ij}\, \bar{\ell}_{L_i}\ell_{R_j} + G_{SL}^{ij}\, \bar{\ell}_{R_i}\ell_{L_j} \right) F_{\mu\nu}F^{\mu\nu} \nonumber \\
	&+ \left( \tilde{G}_{SR}^{ij}\, \bar{\ell}_{L_i}\ell_{R_j} + \tilde{G}_{SL}^{ij}\, \bar{\ell}_{R_i}\ell_{L_j} \right) \tilde{F}_{\mu\nu}F^{\mu\nu} + \text{H.c.}\,,
\end{align}
where \(\ell_i,\,\ell_j = e,\,\mu,\,\tau\) denote the lepton flavors, and \(F_{\mu\nu}\) and \(\tilde{F}_{\mu\nu} = \frac{1}{2}\epsilon_{\mu\nu\sigma\lambda}F^{\sigma\lambda}\) are the electromagnetic field strength tensor and its dual, respectively. The operators involve left- and right-handed chiral components of the leptons, with \(G_{SR}^{ij}\), \(G_{SL}^{ij}\), \(\tilde{G}_{SR}^{ij}\), and \(\tilde{G}_{SL}^{ij}\) parameterizing the interaction strengths. These effective couplings are dimensionful quantities, scaling as \(1/\Lambda^3\), where \(\Lambda\) is the cutoff energy scale of the EFT. This scaling behavior is consistent with the suppression expected for dimension-7 operators in an EFT extension of the SM, where LFV interactions are forbidden at tree level but can arise in new physics models. The dual field strength tensor \(\tilde{F}_{\mu\nu}\) introduces additional complexity to the operator structure, which remains underexplored in the literature, thereby offering the potential for new LFV channels in high-energy experiments.

The local interaction \(\gamma\gamma\,\bar\ell_i\ell_j\) can in principle arise at one loop via the dimension‑5 dipole operator,
\(\bar\ell_i\sigma^{\mu\nu}\ell_j\,F_{\mu\nu}\).
Naïve EFT power counting would then predict that this operator dominates over the dimension‑7 structure 
\(\bar\ell_i\ell_j\,F_{\mu\nu}F^{\mu\nu}\).
However, in well‑motivated UV completions—such as GIM‑suppressed vector‑like lepton loops \cite{Wilczek:1977wb,Bowman:1978kz} or scalar mediators with loop‑induced diphoton form factors \cite{Hisano:2010es}—the dipole can be parametrically suppressed, inverting the hierarchy and rendering the double‑photon operator the leading cLFV effect.

\subsection{EFT parameter space}
Realistic predictions require knowledge of the current experimental bounds on the couplings of interest, |$G_{ij}$|. For the case of |$G_{e\mu}$|, these constraints are comprehensively analyzed in Ref.~\cite{Fortuna:2023paj}, considering processes such as radiative decays $\ell_i \to \ell_j \gamma$, $\ell_i \to \ell_j \gamma\gamma$, and coherent $\ell_i \to \ell_j$ conversion in nuclei. The decay of a charged lepton into three bodies, specifically $\mu^- \to e^- e^+ e^-$, can in principle be used to place constraints on the effective coupling $|G_{e\mu}|$. However, the resulting bound is anticipated to be less stringent than that derived from the radiative decay $\mu \to e \gamma$. This expectation is based on the current experimental upper limits: $\text{BR}(\mu \to 3e) < 1.0 \times 10^{-12}$~\cite{ParticleDataGroup:2024cfk}, which is less severe by approximately an order of magnitude than the bound on the radiative channel, $\text{BR}(\mu \to e \gamma) < 4.2 \times 10^{-13}$~\cite{MEG:2016leq}. Consequently, the process $\mu \to e \gamma$ provides the dominant constraint on the $|G_{e\mu}|$ parameter space.
\subsubsection{$\ell_i\to \ell_j\gamma$ decay}
\begin{figure}[!htb]
	\centering
	\includegraphics[scale=0.2]{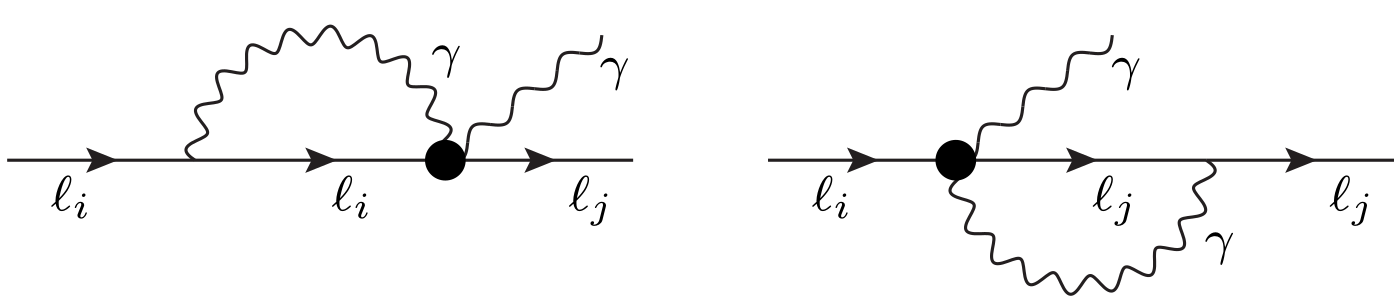}
	\caption{Feynman diagrams that contribute to the $\ell_i\to\ell_j\gamma$ decay from the effective dimension-7 diphoton operators represented by the black circles.}
	\label{FDliligamma}
\end{figure}

At the level of the effective Lagrangian, the decay \(\ell_i \to \ell_j \gamma\) is generated at the one-loop level~\cite{Fortuna:2022sxt}, as shown in Fig.~\ref{FDliligamma}\footnote{There are two additional tadpole diagrams formed by self-contracting the two photons in the effective vertex, with the final-state photon emitted from an external lepton leg. However, since the loop consists of massless photons, the resulting scale-less integrals vanish.}.

 For our numerical analysis, we retain the leading logarithmic contributions to the decay rate. This approximation yields:
\begin{equation}\label{rate_gamma}
	\Gamma(\ell_i \to \ell_j \gamma) \sim \frac{\alpha |G_{ij}|^2}{256\pi^4} m_i^7 \log^2\left(\frac{\Lambda^2}{m_i^2}\right)\,,
\end{equation}
where \(m_i\) is the mass of the decaying lepton, and the final-state lepton mass has been neglected for simplicity.

 The effective coupling \(|G_{ij}|\) encapsulates the relevant operator coefficients and it is defined as:
\begin{equation} \label{Gij}
	|G_{ij}|^2 = |G_{SL}^{\,ij}|^2 + |G_{SR}^{\,ij}|^2 + |\tilde{G}_{SL}^{\,ij}|^2 + |\tilde{G}_{SR}^{\,ij}|^2\,.
\end{equation}

A key feature of this result is the apparent non-decoupling behavior, which manifests as a logarithmic enhancement of the decay rate. However, this embossing is naturally mitigated by the \(1/\Lambda^3\) suppression of the dimensionful couplings \(G_{ij}\), consistent with the scaling behavior expected in an EFT framework. This interplay ensures the theoretical consistency of the EFT description while providing a robust prediction for cLFV observables.

\subsubsection{$\ell_i\to \ell_j\gamma\gamma$ decay}
\begin{figure}[!htb]
	\centering
	\includegraphics[scale=0.4]{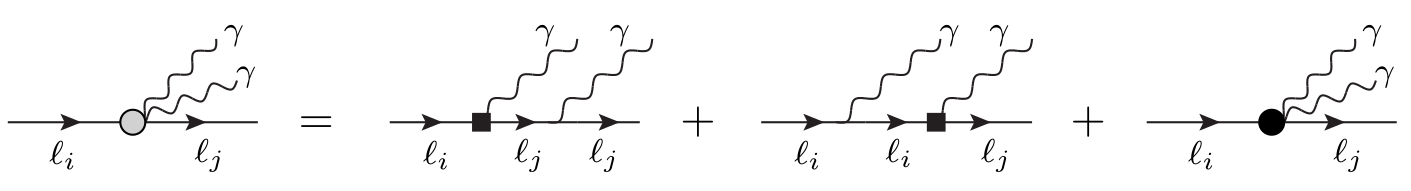}
	\caption{Feynman diagrams that induce the $\ell_i\to\ell_j\gamma\gamma$ decays from the effective dimension-7 diphoton operators (black circle). The black square represents higher order contributions.}
	\label{FDliligammagamma}
\end{figure}

These processes proceed at tree level, a mechanism illustrated in Fig.~\ref{FDliligammagamma} and described by the effective Lagrangian in Eq.~\eqref{Lagrangian_Q}. The decay width follows as 
\begin{equation}\label{liljgg}
	\Gamma(\ell_i \to \ell_j \gamma\gamma) = \frac{\alpha |G_{ij}|^2}{3840\pi^4} m_i^7,
\end{equation}
As evident from Eq.~\eqref{liljgg}, this amplitude exhibits an explicit non-decoupling behavior. Nevertheless, consistent with our earlier discussion, the coupling $|G_{ij}|$ carries an implicit $\Lambda^{-3}$ suppression, which ultimately restores the decoupling property.
\subsubsection{$\ell_i\to \ell_j$ conversion in nuclei}
The dominant contributions to $\mu\to e$ conversion in nuclei can be categorized into two mechanisms, as identified in Ref.~\cite{Davidson:2020ord}. The first originates from the direct interaction of leptons with the static electromagnetic field of the nucleus, mediated by the $e\mu\gamma\gamma$ vertex. The second, a short-distance contribution, stems from the loop-induced mixing of the $e\mu F_{\mu\nu}F^{\mu\nu}$ operator into the scalar proton current $eP_X\mu\bar{p}p$ (with $X=L,R$). For this latter mechanism, the naive loop suppression is overcome by the cumulative enhancement from overlap integrals, energy ratios, and numerical factors. Furthermore, operators involving $F_{\mu\nu}\tilde{F}^{\mu\nu}$, which are proportional to $\vec{E}\cdot\vec{B}$, yield a negligible contribution within the nuclear environment and can therefore be safely disregarded. The $\mu\to e$ conversion process in nuclei has been comprehensively analyzed in Ref.~\cite{Davidson:2020ord}. For $\ell\to\tau$ transitions, we refer to the computations in Ref.~\cite{Fortuna:2023paj}. 

\begin{figure}[!htb]
	\centering
	\includegraphics[scale=0.2]{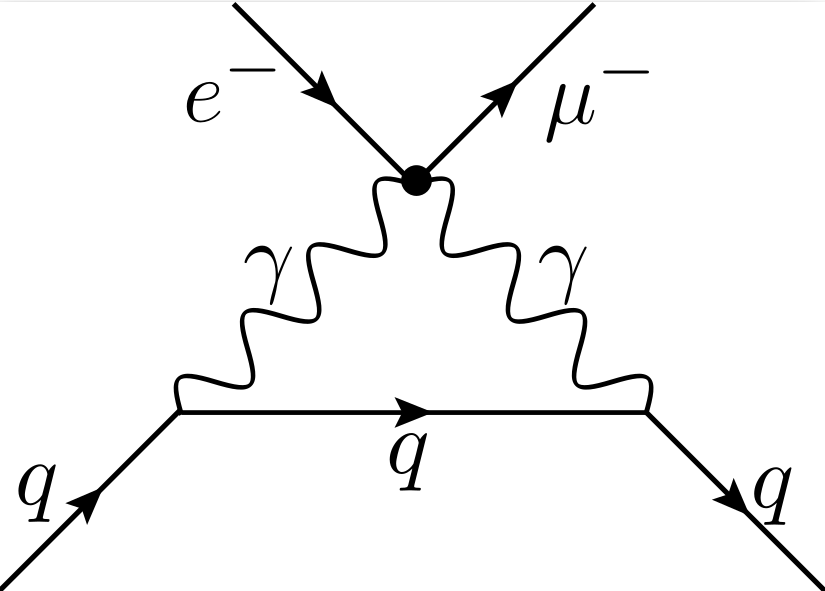}
	\caption{Feynman diagram that contribute to the $e\to\mu$ conversion from the effective dimension-7 diphoton operators represented by the black circle.}
	\label{FDconversion}
\end{figure}

In light of our focus on the $e\mu\gamma\gamma$ interaction (see Fig.~\ref{FDconversion}), we provide only the relevant conversion rate for a nucleus $A$, as derived in Ref.~\cite{Davidson:2020ord}.

\begin{equation}
	\mu A\to e A = \frac{4m^4_\mu}{\Gamma_{\rm cap}}|G_{e\mu}|^2 \Big|m_\mu F_A+\frac{18\alpha m_p}{\pi}S_A^{(p)}\Big|^2,
\end{equation} 
where $\Gamma_{\rm cap}$ is the muon capture rate on nucleus $A$ and $F_A/S_A^{(p)}$ are overlap integrals that can be found in Refs.~\cite{Davidson:2020ord} and \cite{Kitano:2002mt}, respectively.

As shown in \cite{Fortuna:2022sxt}, the most stringent constraint comes from the upper limit on $\ell_i\to \ell_j\gamma$. Then, from the decay rate in Eq.~(\ref{rate_gamma}) and the experimental upper limits on ${\rm BR}(\ell_i \to \ell_j \gamma)$~\cite{MEG:2016leq, BaBar:2009hkt, Belle:2021ysv},
the following constraints on the effective couplings $|G_{ij}|$ are found:  
\begin{align}\label{ul_Gij}
	|G_{\mu e}|&\lesssim 9.5\times10^{-10}\left(1+0.15\ln{\frac{\Lambda}{100~{\rm GeV}}}\right)^{-1}~{\rm GeV}^{-3}\,,\\
	|G_{\tau e}|&\lesssim 8.4\times10^{-9}\left(1+0.25\ln{\frac{\Lambda}{100~{\rm GeV}}}\right)^{-1}~{\rm GeV}^{-3}\,,\nonumber \\
	|G_{\tau \mu}|&\lesssim 9.5\times10^{-9}\left(1+0.25\ln{\frac{\Lambda}{100~{\rm GeV}}}\right)^{-1}~{\rm GeV}^{-3}\,,\nonumber
\end{align} 
where the bounds exhibit a mild logarithmic dependence on the cutoff scale $\Lambda$. To provide a complete overview, Tab.~\ref{tab:BRsExp} lists the existing experimental bounds on the aforementioned processes.

\begin{table}[t!]
	\begin{center}
		\setlength{\tabcolsep}{6pt}
		\begin{tabular}{lllll}
			\hline
			\hline
			cLFV obs. & \multicolumn{2}{l}{Current upper limit  (90\%CL)} & \multicolumn{2}{l}{Expected future limits }\\
			\hline
			$\mu\to e\gamma$ &  $4.2\times10^{-13}$ & MEG (2016)~\cite{MEG:2016leq}& $6\times10^{-14}$ & MEG-II~\cite{MEGII:2018kmf} \\
			$\mu\to e\gamma\gamma$ &  $7.2\times10^{-11}$ & Crystal Box (1986)~\cite{Grosnick:1986pr}& --- & --- \\
			$\mu A\to eA$ &  $7\times10^{-13}$ & Sindrum II (2006)~\cite{SINDRUMII:2006dvw}& $6.2\times10^{-16}$ & Mu2e~\cite{Mu2e:2022ggl}\\
			$\tau\to e\gamma$ &  $3.3\times10^{-8}$ &BaBar (2010)~\cite{BaBar:2009hkt}& $9\times10^{-9}$ & Belle-II~\cite{Banerjee:2022xuw} \\
			$\tau\to e\gamma\gamma$ &  $2.5\times10^{-4}$ & Bryman {\it et al.} (2021)~\cite{Bryman:2021ilc}& --- &--- \\
			$\tau\to \mu\gamma$ &  $4. 2\times10^{-8}$ &Belle (2021)~\cite{Belle:2021ysv}& $6.9\times10^{-9}$& Belle-II~\cite{Banerjee:2022xuw} \\
			$\tau\to \mu\gamma\gamma$ &  $1.5\times10^{-4}$ & ATLAS (2017)~\cite{Angelozzi:2017oeg}& --- &--- \\
			$\ell \mathcal{N}\to \tau X$ & 
			--- & --- & $[10^{-13},10^{-12}]$ & NA64~\cite{Gninenko:2018num} \\
			\hline
			\hline
		\end{tabular}
		\caption{Experimental upper bounds and future expected sensitivities for the set of cLFV transitions relevant to our analysis. In the process $\mu A\to eA$ conversion, $A=Au$ for the current upper limit,  while $A=Al$ for the expected future limit. 
			Note there are also promising sensitivities at STCF for LFV $\tau$ decays~\cite{Achasov:2023gey}, nevertheless, we will mostly use Belle-II numbers for the future, as they are expected to be released sooner.} \label{tab:BRsExp}  
	\end{center}
\end{table}

We now introduce nonchiral coefficients, which are more convenient for Monte Carlo simulations. The scalar and pseudoscalar coefficients are defined as:
\begin{align}\label{couplings_SP}
	G_S^{ij} &= \frac{G_{SR}^{ij} + G_{SL}^{ij}}{2}, \quad G_P^{ij} = \frac{G_{SR}^{ij} - G_{SL}^{ij}}{2}, \nonumber \\
	\tilde{G}_S^{ij} &= \frac{\tilde{G}_{SR}^{ij} + \tilde{G}_{SL}^{ij}}{2}, \quad \tilde{G}_P^{ij} = \frac{\tilde{G}_{SR}^{ij} - \tilde{G}_{SL}^{ij}}{2}\,,
\end{align}
which leads us to derive the following constraint:
\begin{equation}\label{nonchiralcoupling}
	|G_{ij}|^2=2\big(|G_P^{ij}|^2+|G_S^{ij}|^2+|\tilde{G}_P^{ij}|^2+|\tilde{G}_S^{ij}|^2\big)\,.
\end{equation}

\begin{figure}[!htb]
	\centering
    \includegraphics[scale=0.2]{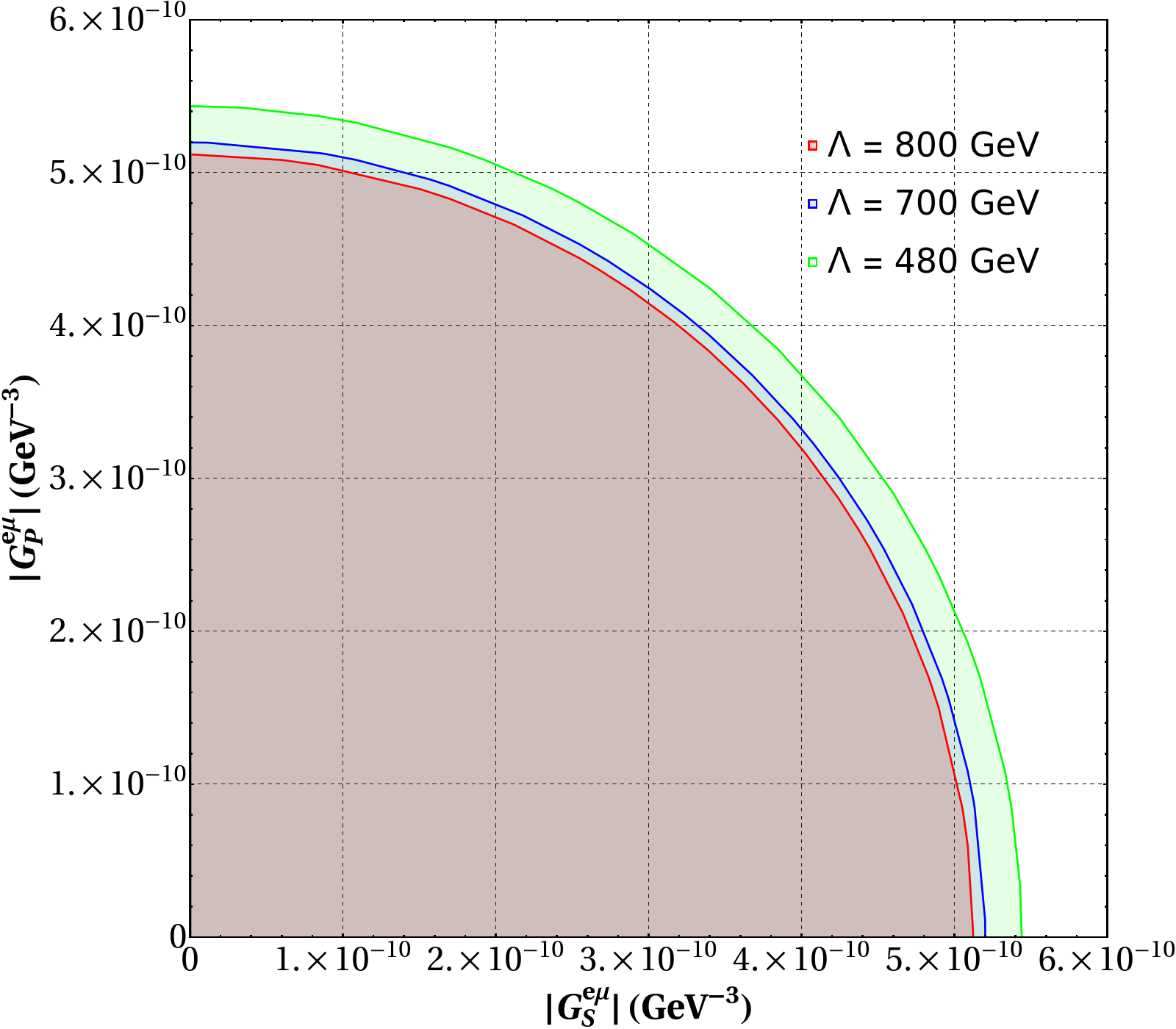}
    \caption{Parameter space allowed by the upper limit on $\rm {BR}(\mu\to e\gamma)$ in the $|G_S^{e\mu}|-|G_P^{e\mu}|$ plane. The green, blue and red regions correspond to cut-off scales of $\Lambda=480,\,700,\,800$ GeV, respectively.}
    \label{ParamSpace1}
   \end{figure}

From Eqs.~\eqref{ul_Gij} and \eqref{nonchiralcoupling}, we obtain the parameter space in the $|G_S^{e\mu}|-|G_P^{e\mu}|$ plane, which is presented in Fig.~\ref{ParamSpace1} \footnote{Since the contribution of the proportional term to $\tilde{F}_{\mu\nu} F^{\mu\nu}$ is subdominant --up to five orders of magnitude smaller than ${F}_{\mu\nu} F^{\mu\nu}$-- the couplings $\tilde{G}^{ij}_{S(P)}$ do not have a significant impact for the study addressed in this work.}. The shaded regions represent the parameter space allowed by the current upper limit on $\rm {BR}(\mu\to e\gamma)$. The red, blue, and green areas correspond to the allowed regions for cutoff scales of $\Lambda = 800 \text{ GeV}$, $\Lambda = 700 \text{ GeV}$, and $\Lambda = 480 \text{ GeV}$, respectively. We find that the $|G_S^{e\mu}|$ and $|G_P^{e\mu}|$ couplings are stringently constrained to values of order $\lesssim \mathcal{O}(10^{-10})$, with slightly larger allowed parameter spaces emerging for lower values of $\Lambda$. Thus, if one considers \(\Lambda \gtrsim 2\,\sqrt{s},\) then contributions from a dimension‑\(d\) operator are guaranteed to be suppressed by at least
	\((\frac{\sqrt{s}}{\Lambda})^{d-4},\) thereby maintaining the consistency of the EFT expansion and keeping non‑renormalizable terms under control. For the projected \(e^+e^-\) runs at \(\sqrt{s}=240\) GeV and \(350\) GeV, this implies \(\Lambda \gtrsim 480~\text{GeV}\;\text{and}\;\Lambda \gtrsim 700~\text{GeV},\) values which coincide with our benchmark cutoffs, ensuring that all three BMPs remain within a parametrically valid regime.\\

\section{Collider analysis}\label{Collider}
\begin{figure}[!htb]
	\centering
	\includegraphics[scale=0.2]{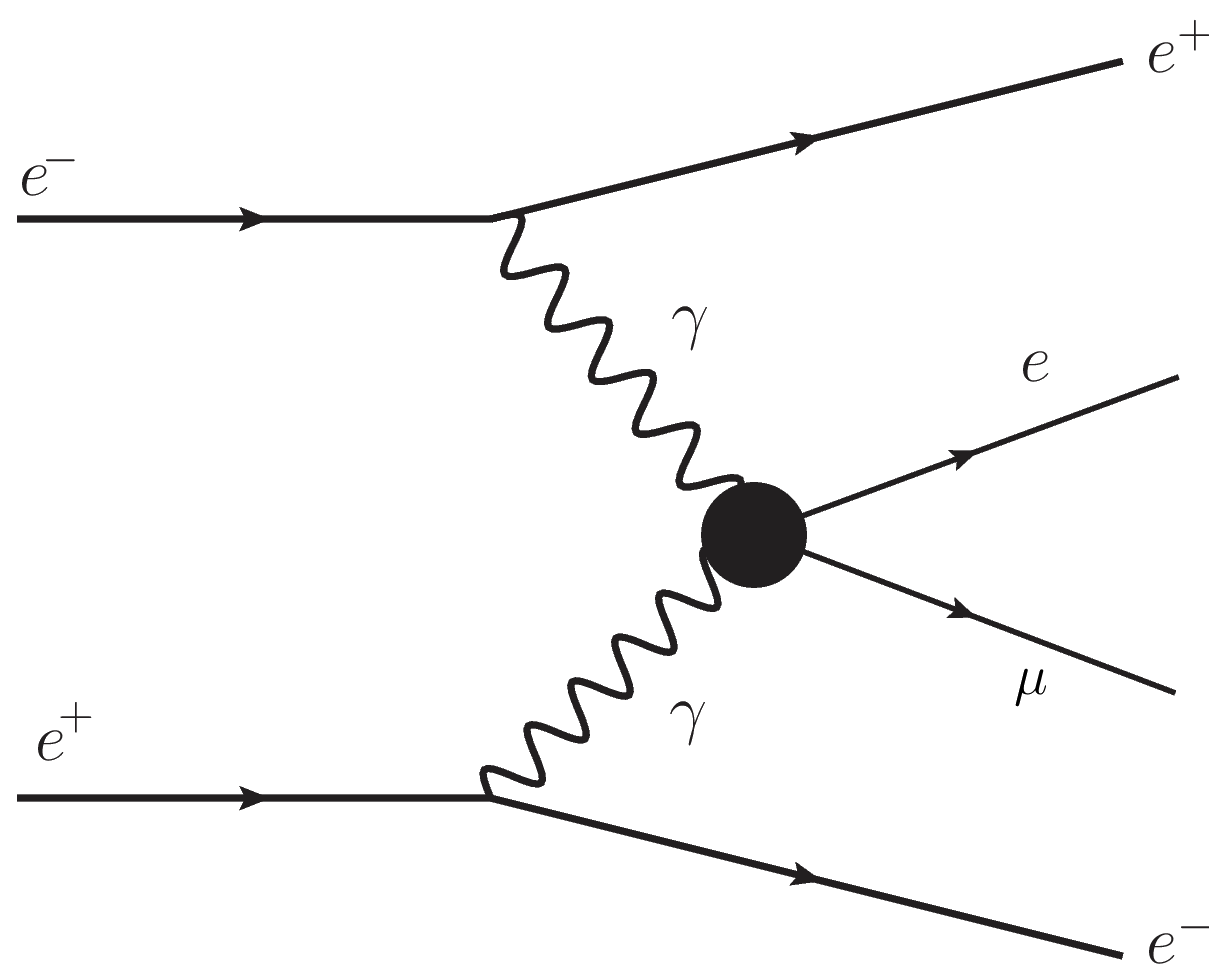}
	\caption{Feynman diagram of the signal $e^-e^+\to e^- e^+ e\,\mu$. The black circle represents the effective interaction.} 
	\label{FDsignal}
\end{figure}

In this section, we present a Monte Carlo simulation of the signal process, $e^- e^+ \to e^- e^+ e \mu$, as depicted in Fig.~\ref{FDsignal}, along with the relevant Standard Model backgrounds. We also outline the analysis strategy employed to separate the signal from background contributions.

\subsection{Signal and background}
\begin{itemize}
	\item \textbf{\textit{Signal:}} We study the final state $e^-e^+ e\,\mu$, where $e^-$ and $e^+$ originate from the initial-state scattered pair. These are expected to exhibit higher transverse momentum compared to the charged particles ($e\mu$) produced via photon fusion. The analysis of the full process is essential, as the kinematic properties of the initial-state $e^-e^+$ pair provide critical information for distinguishing the signal from background. In particular, the transverse momentum of the scattered electron serves as a powerful discriminant, significantly reducing background contamination.\\    
	To illustrate the phenomenological behavior, we estimate the order of magnitude for the effective cross-section and the expected event yield by defining three benchmark points (BMPs) as follows
	\begin{itemize}	
		\item {\bf BMP1}: $|G_S^{e\mu}|=4\times 10^{-10}$ GeV$^{-3}$,  $|G_P^{e\mu}| = 1\times 10^{-10}$ GeV$^{-3}$,  $|\tilde{G}_S^{e\mu}|=1\times 10^{-13}$ GeV$^{-3}$,  $|\tilde{G}_P^{e\mu}|=1\times 10^{-10}$ GeV$^{-3}$,  $\Lambda=800$ GeV,
		\item {\bf BMP2}: $|G_S^{e\mu}|=9\times 10^{-11}$ GeV$^{-3}$,  $|G_P^{e\mu}| = 5\times 10^{-10}$ GeV$^{-3}$,  $|\tilde{G}_S^{e\mu}|=|\tilde{G}_P^{e\mu}|=0$,  $\Lambda=700$ GeV,
		\item {\bf BMP3}: $|G_S^{e\mu}|=2\times 10^{-10}$, $|G_P^{e\mu}|=1.65\times 10^{-10}$ GeV$^{-3}$,  $|\tilde{G}_S^{e\mu}|=2\times 10^{-10}$ GeV$^{-3}$ GeV$^{-3}$,  $|\tilde{G}_P^{e\mu}|=1.85\times 10^{-11}$ GeV$^{-3}$,  $\Lambda=480$ GeV.
	\end{itemize}
   Although the couplings $G_{S\,(P)}^{e\mu}$ and $\tilde{G}_{S\,(P)}^{e\mu}$ are suppressed, the signal cross section grows at higher center-of-mass energies, as illustrated in Fig.~\ref{XSsignal}. Moreover, both colliders are projected to achieve signal event rates comparable to—or even exceeding—those of the dominant background processes.
   
   \begin{figure}[!htb]
   	\centering
   	\includegraphics[scale=0.6]{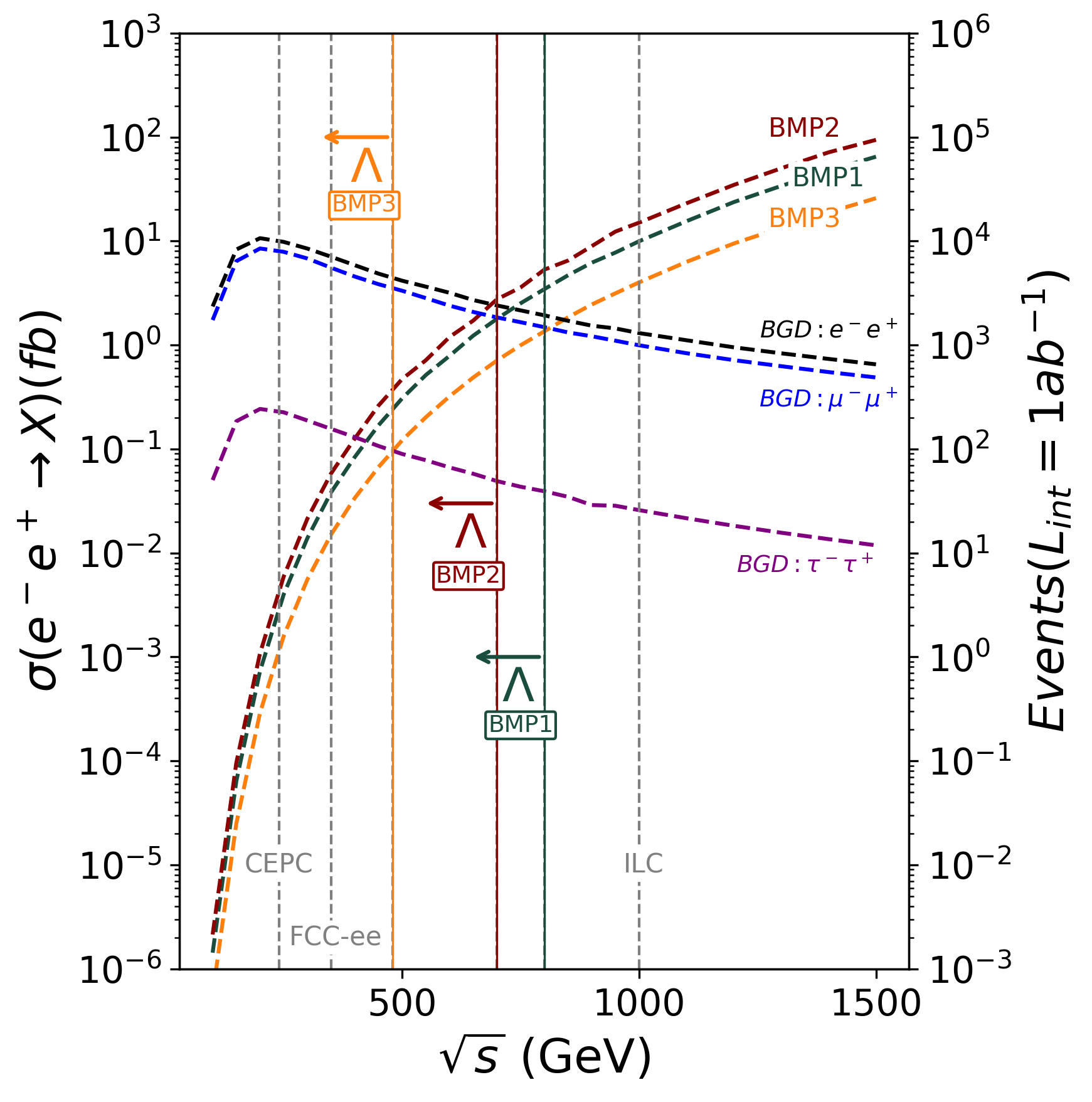}
   	\caption{Production cross sections of the signal for the three BMPs and the dominant SM background. The number of events expected for an integrated luminosity of $\mathcal{L}_{\rm int}=1$ ab$^{-1}$ is shown on the right vertical axis. The dashed vertical lines correspond to center-of-mass energies associated with the next generation of $e^-e^+$ colliders. Finally, the solid vertical lines represent the cutoff scales: $\Lambda_{\text{BMP1}}=800$ GeV (green line), $\Lambda_{\text{BMP2}}=700$ GeV (wine color line), $\Lambda_{\text{BMP3}}=480$ GeV (orange line). } 
   	\label{XSsignal}
   \end{figure}
   \paragraph{Choice of the $e\mu$ final state}
   While the $\gamma\gamma\to \tau e$ and $\gamma\gamma\to \tau\mu$ channels could in principle benefit from potentially larger LFV couplings, their experimental signature is substantially compromised by the neutrinos from $\tau$ decays. These introduce significant missing transverse energy ($\slashed{E}_T$) from undetected neutrinos, degrading the kinematic reconstruction crucial for background suppression. In contrast, the $e\mu$ final state offers a direct, fully reconstructible signature with minimal $\slashed{E}_T$, providing superior sensitivity despite the more constrained $G_{e\mu}$ coupling. This advantage is particularly pronounced in the clean environment of $e^+e^-$ colliders, where the $e\mu$ channel allows for sharper discrimination against dominant QED backgrounds.

	\item \textbf{\textit{Background}}: Thanks to the clean environment of $e^-e^+$ collisions, the dominant SM backgrounds arise from 
    \begin{itemize}
        \item $e^-e^+\to e^-e^+e^-e^+$,
        \item $e^-e^+\to e^-e^+\mu^-\mu^+$,
        \item $e^-e^+\to e^-e^+ \tau\tau \to e^-e^+ e\mu +\slashed{E}_T$.
    \end{itemize}

\textit{Misidentified Leptons ("Fake Leptons"):} A significant background can arise from jets fragmenting into hadrons that are mis-reconstructed as leptons. These processes are given as follows

      \begin{itemize}

          \item $e^-e^+\to e^-e^+ \pi^-\pi^+$,
          \item $e^-e^+\to e^-e^+ \pi^-\pi^+ +\slashed{E}_T$.
          \\
In \(e^{+}e^{-}\) collisions, the pion-to-electron misidentification rate---where charged pions (\(\pi^{\pm}\)) are falsely reconstructed as electrons---typically ranges from $0.1\%$ to $1\%$ for momentum in the \(\sim 1\text{--}10\ \text{GeV}\) range, depending on the selection criteria applied. In contrast, the corresponding misidentification rate for muons is substantially lower, at the level of $0.01\%$ to $0.1\%$. As a result, the contribution of this background to the final signal region is generally small, especially in analyses that require well-identified, prompt leptons.
                    \end{itemize}
             \textit{ Radiative Processes with Photon Conversions:}
          \begin{itemize}
          	\item  These processes involve a final state with photons that convert into a dilepton pair, $e^-e^+\to\gamma\gamma$ (with $\gamma\to \ell^-\ell^+$), can produce two non-prompt low-mass leptons. However, due to their markedly different kinematic properties compared to the signal, these processes do not constitute a dominant background. Events containing a reconstructed secondary vertex consistent with photon conversion are vetoed, ensuring a negligible impact on the signal-to-background discrimination.
          \end{itemize}
                          \end{itemize}

\begin{table}[b!]
	\centering
	\caption{Production cross sections of the signal $e^- e^+\to e^- e^+ e\, \mu$ for the three BMPs. *Note that BMP3, with $\Lambda = 480$ GeV, is only valid for $\sqrt{s}=240$ GeV, since the condition $\Lambda \gtrsim 2\sqrt{s}$ required for EFT consistency is not satisfy at $\sqrt{s}=350$ GeV. This explains the \textit{No valid} entry for BMP3 at $\sqrt{s}=350$ GeV.}
	\label{tb:XS-signal}
	
	\renewcommand{\arraystretch}{1.2} 
	\setlength{\tabcolsep}{6pt} 
	
	\begin{tabular}{|c|c|c|}
		\hline
		BMPs & $\sqrt{s}=240$ GeV & $\sqrt{s}=350$ GeV \\
		\hline 
		\hline
		BMP1 & $0.003$ fb & $0.038$ fb \\
		\hline
		BMP2 & $0.005$ fb& $0.120$ fb \\
		\hline
		BMP3 & $0.0012$ fb & \textit{No valid}*  \\
		\hline 
	\end{tabular}
\end{table}
\begin{table}[b!]
	\caption{Cross sections of the dominant SM background.}\label{tb:XS-BGD}
	\begin{centering}
		\begin{tabular}{|c|c|c|}
			\hline 
			Energy & $\sigma(e^{-}e^{+}\to e^{-}e^{+}e^{-}e^{+})$ & $\sigma(e^{-}e^{+}\to e^{-}e^{+}\mu^{-}\mu^{+})$\tabularnewline
			\hline 
			\hline 
			240 GeV & 10.2 fb & 7.6 fb\tabularnewline
			\hline
			350 GeV & 7.4 fb & 5.6 fb
			\tabularnewline
			\hline
		\end{tabular}
		\par\end{centering}
\end{table}
The numerical cross sections for the signal and dominant background processes are presented in Tables~\ref{tb:XS-signal} and \ref{tb:XS-BGD}, respectively. Meanwhile, Fig.~\ref{XSsignal} provides an overview of the general behavior of the cross section as a function of the center-of-mass energy. The number of signal and background events produced is shown on the right vertical axis, assuming an integrated luminosity of 1 ab$^{-1}$. The vertical lines within the graph represent the different \(e^- e^+\) colliders at their maximum projected center-of-mass energies\footnote{CLIC -not shown- will search up to $\sqrt{s}=3$ TeV.}. The cutoff scales corresponding to our BMPs imply that the signal cross section surpasses the background at \(\sqrt{s}\approx700\) GeV for BMP2, 750 GeV for BMP1, and 850 GeV for BMP3. However, because these energies coincide with or exceed the respective EFT cutoff \(\Lambda\), the effective‑field‑theory description may become unreliable, indicating that a dedicated UV‑complete or Higgs-EFT analysis would be required to study these regimes with confidence.



Regarding our computational framework, we first implement the relevant interactions using \texttt{LanHEP} ~\cite{Semenov:2014rea} for compatibility with \texttt{MadGraph5}~\cite{Alwall_2011}. The generated events are then interfaced with \texttt{Pythia8}~\cite{Sjostrand:2014zea} for parton showering and hadronization, followed by detector simulation using \texttt{Delphes3}~\cite{deFavereau:2013fsa}. For this purpose, we employ the \texttt{delphes\_card\_CEPC.tcl}~\cite{DelphesCEPC} and \texttt{delphes\_card\_IDEA.tcl}~\cite{DelphesIDEA} detector configurations for the CEPC and FCC-ee, respectively.

Following a kinematic analysis using traditional cut-based selections in \texttt{MadAnalysis5}~\cite{Conte:2012fm}, we observed that most discriminating observables provided limited separation power between signal and background. To enhance the sensitivity, we employed Multivariate Analysis (MVA) techniques, combining these observables into a more powerful classifier. For the MVA training, we utilized a Neural Network (NN) approach \cite{lecun1998gradient, rumelhart1986learning, vaswani2017attention}\footnote{The datasets and analysis code are available upon request.}, with input features derived from the kinematics of the final-state particles: the scattered electron and positron, and the centrally produced electron and muon pair. The kinematic variables include the transverse momentum $p_T$, pseudo-rapidity $\eta$, the azimuthal angle $\phi$. The signal and background samples are scaled to the expected number of candidates, which is calculated based on the integrated luminosity and cross-sections. The NN selection is optimized individually for each channel to maximize the figure of merit, i.e., the signal significance, defined as $\displaystyle \mathcal{S} = \frac{N_S}{\sqrt{N_S + N_B + (\kappa \cdot N_B)^2}}$, where \(N_S\) and \(N_B\) denote the number of signal and background events, respectively, and $\kappa$ represents the systematic uncertainty. We assume a total systematic uncertainty combining contributions from luminosity calibration ($1\%$), and background modeling ($3\%$), and it is given by~\cite{CEPCStudyGroup:2023quu}
	\begin{equation}
	    \kappa=\sqrt{(1\%_{\rm lumi})^2+(3\%_{\rm modeling})^2}
		\approx 3\%.
	\end{equation}
		
\begin{figure}[t]
	\centering
    \begin{subfigure}[t]{0.45\textwidth}
    \centering
    \includegraphics[width=\textwidth]{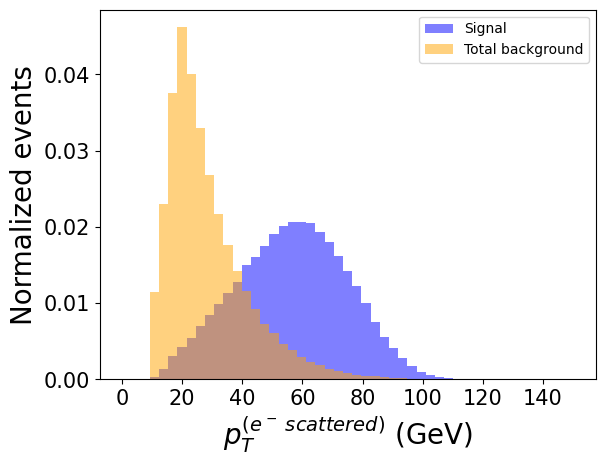}
    \caption{Transverse momentum of the scattered electron.}
    \label{subf:a}
    \end{subfigure}
    \begin{subfigure}[t]{0.46\textwidth}
    \centering
    \includegraphics[width=\textwidth]{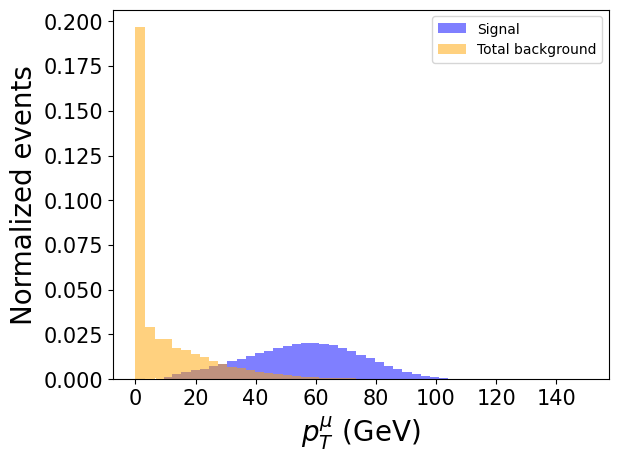}
    \caption{Transverse momentum of the muon.}
    \label{subf:b}
    \end{subfigure}
    \caption{ Normalized distributions for BMP1, (a) \( p_T^{e^-\text{scattered}} \) and the muon (b) \( p_T^{\mu} \). These variables were selected for the multivariate analysis due to their clear discriminating power between the signal and background processes. The number of events for both signal and background has been normalized such that the area under each histogram is unity.} 
\label{fig:distributions}
\end{figure}

Based on the NN performance, the most significant observables for distinguishing the signal from background at $\sqrt{s}=240$ GeV in both CEPC and FCC-ee are: the transverse momentum of the scattered electron $p_T^{(e^-\,\text{scattered})}$, the transverse momentum of the central muon $p_T^{\mu}$, the pseudorapidity of the muon $\eta(\mu)$, and the angular separation between the central muon and the scattered positron, $\Delta R(\mu, e^{+\rm \text{scattered}})$. The distributions of $p_T^{(e^-\text{scattered})}$ and $p_T^{\mu}$ for BMP1 are presented in Figure~\ref{fig:distributions}, demonstrating their discriminating power between signal and background processes.  
The signal's high transverse momentum $p_T^{(e^-\rm {scattered})}$ and $p_T^{\mu}$ spectra stem from the contact-like $\gamma\gamma e\mu$ dimension-7 operator, which couples to photons with large virtuality $Q^2$ and thus enhances high-transverse-momentum configurations. In contrast, the dominant QED backgrounds proceed via quasi-real photon exchange, described by the Equivalent Photon Approximation, which strongly suppresses large $Q^2$ and hence yields softer $p_T$  distributions. This kinematic distinction enables efficient signal-background separation using just a handful of key observables that involve those associated with the final-state $e\mu$ pair from $\gamma\gamma$ interactions, and the kinematic properties of the scattered $e^{\pm}$ from the initial process.  This highlights the importance of analyzing the complete process to achieve signal significances at the evidence/detection level. 

\begin{figure}[t!]
	\centering
		\includegraphics[scale=0.3]{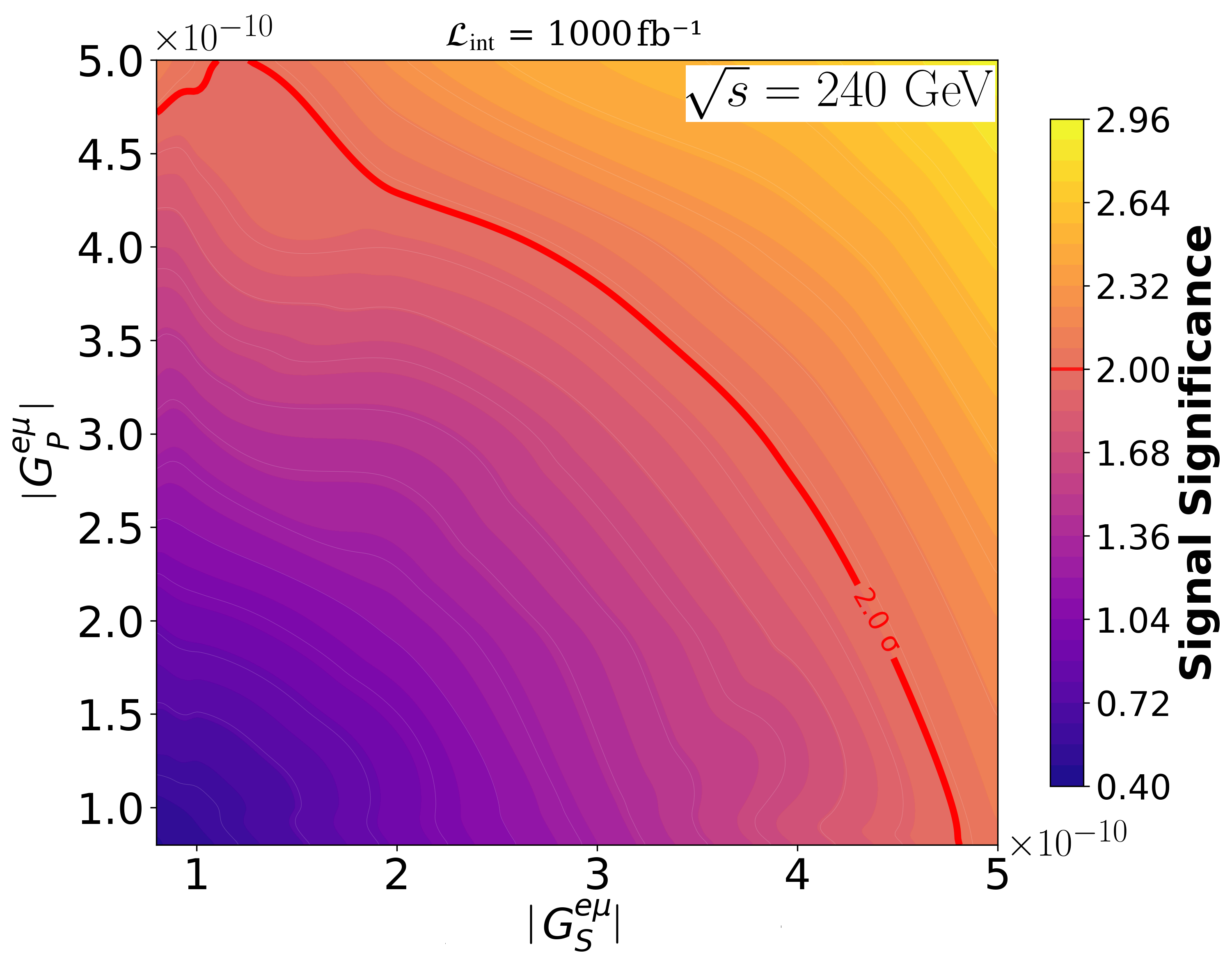}
		\includegraphics[scale=0.3]{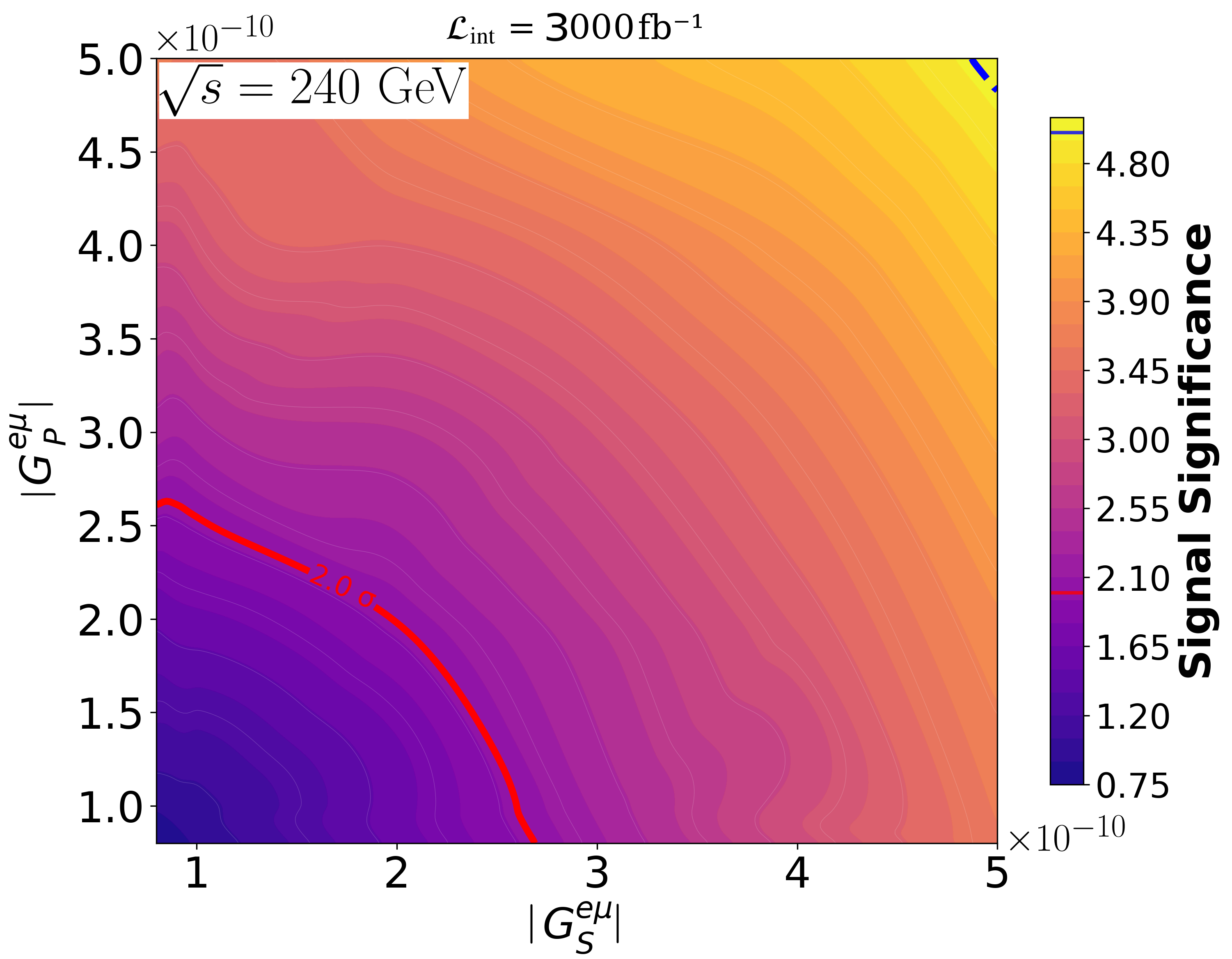}
		\includegraphics[scale=0.3]{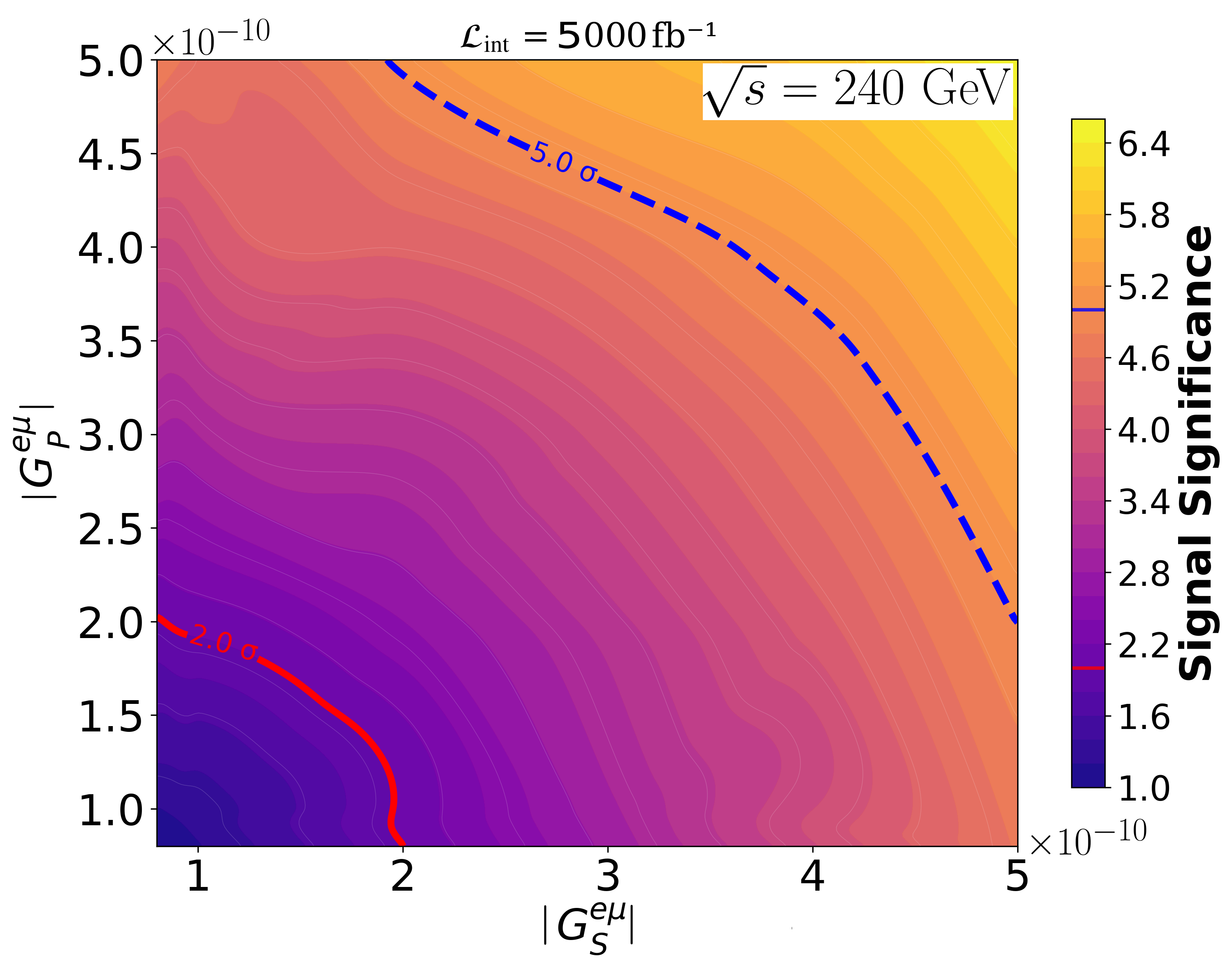}
			\caption{Signal significance as a function of the $|G^{e\mu}_S|$ and $|G^{e\mu}_P|$ couplings for $\mathcal{L}_{\rm int}=1000,\,3000,\,5000$ fb$^{-1}$, corresponding to CEPC at $\sqrt{s}=240$ GeV. Systematic uncertainties ($\kappa\approx3\%$) are included in the significance calculation via Gaussian smearing of efficiencies.}  
	\label{fig:significance240}
\end{figure}

\begin{figure}[t!]
	\centering
	\includegraphics[scale=0.5]{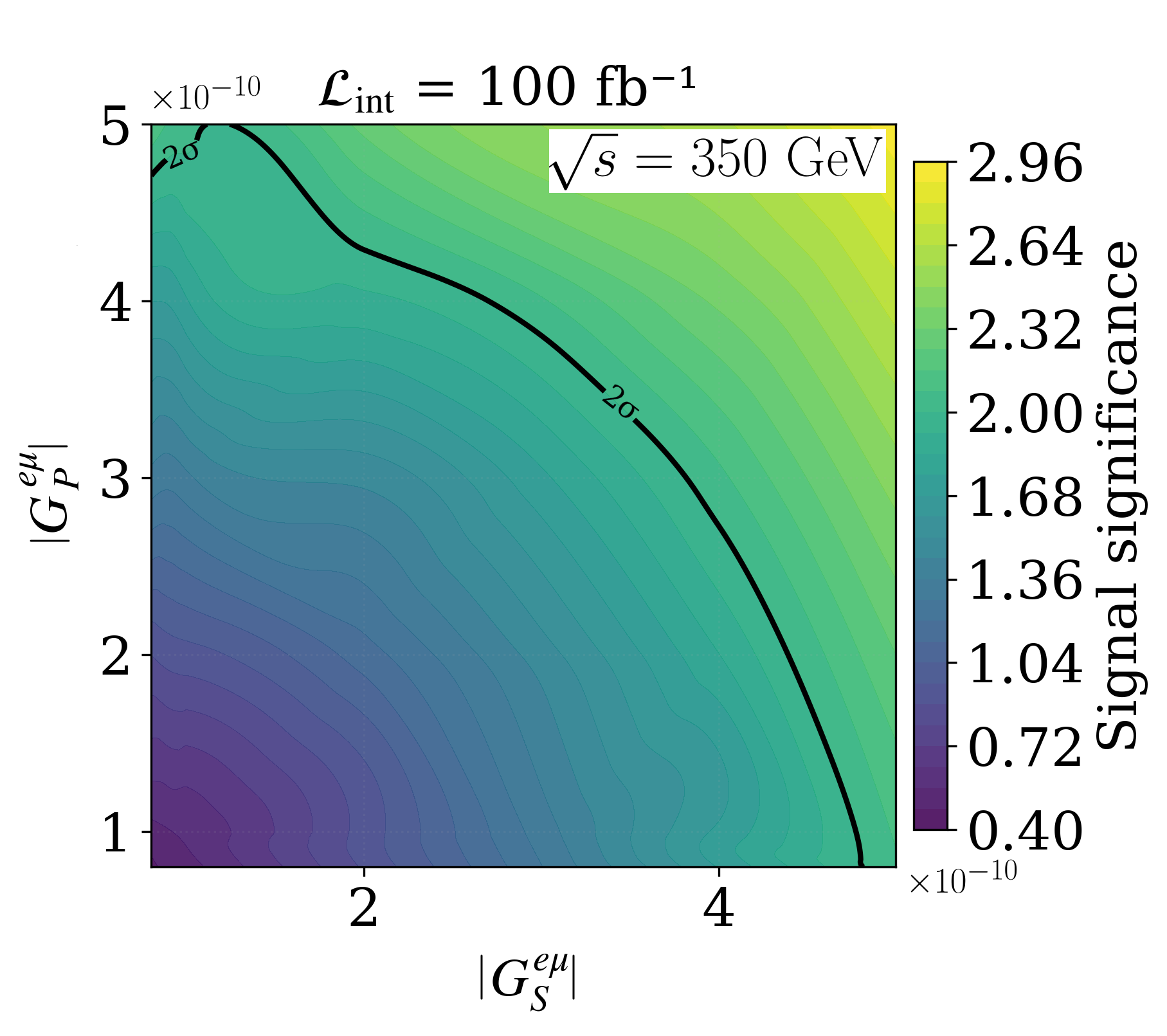}
	\includegraphics[scale=0.5]{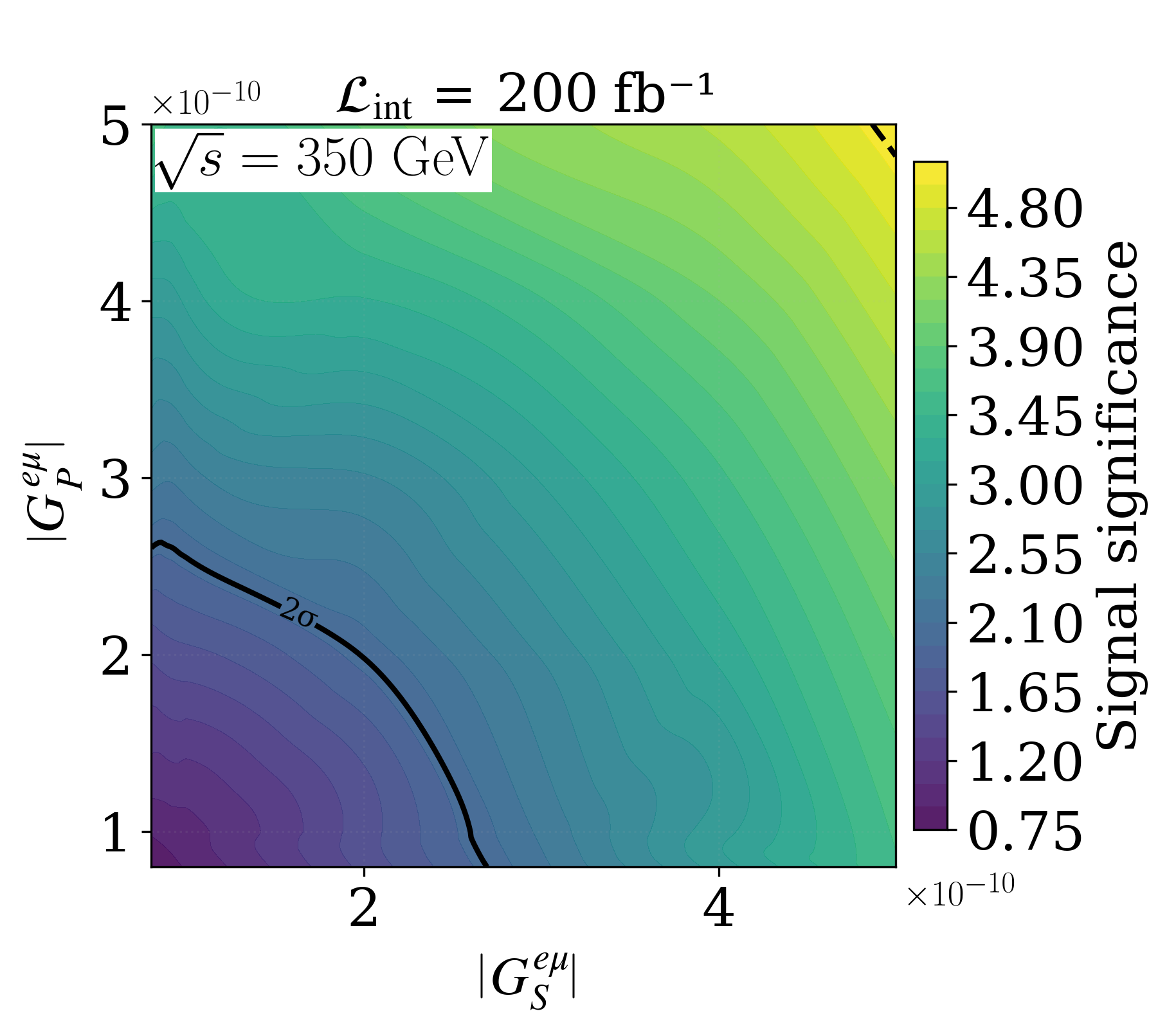}
	\includegraphics[scale=0.5]{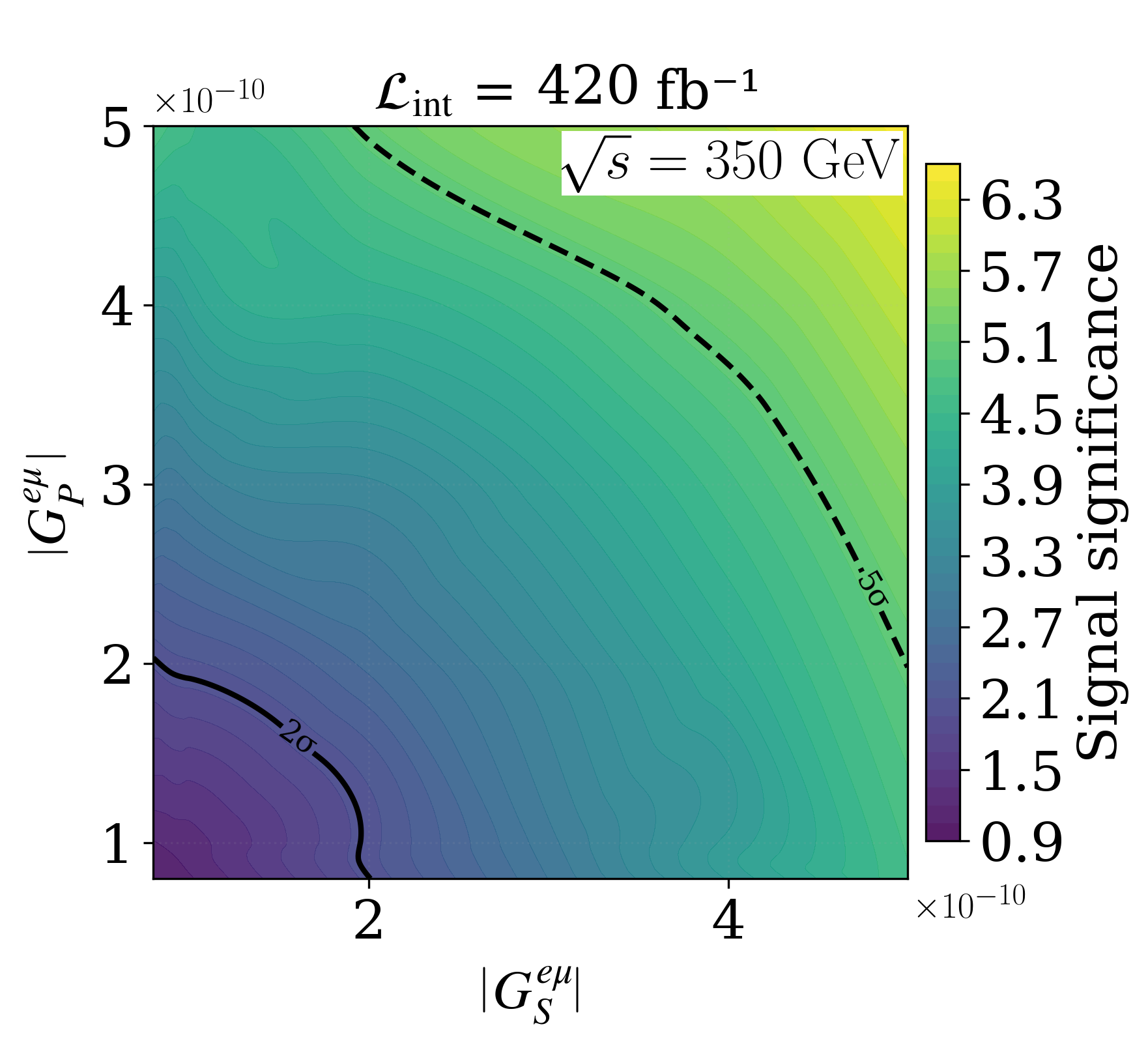}
	\caption{Signal significance as a function of the $|G^{e\mu}_S|$ and $|G^{e\mu}_P|$ couplings for $\mathcal{L}_{\rm int}=100,\,200,\,420$ fb$^{-1}$, corresponding to FCC-ee at $\sqrt{s}=350$ GeV. Systematic uncertainties ($\kappa\approx3\%$) are included in the significance calculation via Gaussian smearing of efficiencies.}  
	\label{fig:significance350}
\end{figure}

Thus, we present in Fig.~\ref{fig:significance240} the projected signal significance $\mathcal{S}$ as a function of the scalar and pseudoscalar effective couplings, $|G^{e\mu}_S|$ and $|G^{e\mu}_P|$, for three integrated luminosities: $\mathcal{L}_{\rm int}=1000,\,3000,\,5000~\text{fb}^{-1}$. These contours correspond to the sensitivity attainable at the CEPC operating at $\sqrt{s}=240~\text{GeV}$. Complementarily, Fig.~\ref{fig:significance350} presents the analogous significance projections for the FCC-ee at a center-of-mass energy of $\sqrt{s}=350~\text{GeV}$. Collectively, these results underscore the outstanding discovery reach of future circular colliders in probing the $\gamma\gamma e\mu$ effective vertex.

The analysis reveals that the possible experimental detection depends on both the accumulated integrated luminosity and the specific values of the $|G^{e\mu}_S|$ and $|G^{e\mu}_P|$ couplings. For the CEPC at $\sqrt{s}=240~\text{GeV}$ with $\mathcal{L}_{\rm int}=5000~\text{fb}^{-1}$, a $5\sigma$ discovery would be achievable for coupling values lying below the blue segmented contour, whereas the red solid contour demarcates the $2\sigma$ exclusion region. Notably, the required integrated luminosity for a given sensitivity is substantially reduced at the FCC-ee  at $\sqrt{s}=350$ GeV, owing to signal cross sections that can be up to an order of magnitude larger than those at $\sqrt{s}=240$ GeV, as illustrated in Fig.~\ref{XSsignal}.

\section{Conclusions}\label{Conclusions}
In this work, we have explored the potential of future circular $e^+e^-$ colliders—specifically, CEPC at $\sqrt{s}=240\ \text{GeV}$ and FCC-ee at $\sqrt{s}=240, 350\ \text{GeV}$—to probe the Lepton Flavor Violating $\gamma\gamma e\mu$ interaction via the process $e^+ e^- \to e^+ e^- e\mu$.

Adopting an effective field theory (EFT) framework consistent with current theoretical and experimental constraints, we demonstrate that, with an integrated luminosity of $3000\ \text{fb}^{-1}$ and couplings of order $|G^{e\mu}_S|\approx |G^{e\mu}_P|\approx 5\times 10^{-10}$, a $5\sigma$ observation is achievable at CEPC operating at $\sqrt{s}=240\ \text{GeV}$. For FCC-ee at $\sqrt{s}=350\ \text{GeV}$, a similar discovery sensitivity can be reached with significantly lower luminosity—$200\ \text{fb}^{-1}$—highlighting the advantage of its high-purity collision environment.

Although future linear colliders such as the ILC and CLIC target higher center-of-mass energies (up to $1$–$3\ \text{TeV}$), the sensitivity in our analysis depends more critically on integrated luminosity than on collision energy. This is due to the theoretical requirement $\Lambda \gtrsim 2\sqrt{s}$, which ensures the validity of the EFT description and imposes a stringent upper bound on the accessible $\sqrt{s}$.

These results position CEPC and FCC-ee as premier facilities for probing charged LFV via photon fusion, offering a powerful complement to direct searches at the LHC and to low-energy rare-decay experiments. Future studies should address systematic uncertainties related to background modeling and detector effects to further refine the projected sensitivities. Moreover, both colliders can produce the signal copiously by leveraging either higher energy or increased luminosity, potentially enabling the rediscovery of the $\gamma\gamma e\mu$ LFV process within their own datasets.

We strongly encourage the experimental collaborations to investigate this signal in the early phases of data collection.
 
\section*{Acknowledgments}

Marcela Marín acknowledges support from the UNAM Postdoctoral Program (POSDOC). We are grateful to Michel Hernández Villanueva, Pablo Roig, and Fabiola Fortuna for their invaluable comments. The work of Marco A. Arroyo-Ureña is supported by the ``Estancias Posdoctorales por México (SECIHTI)'' program and the ``Sistema Nacional de Investigadores e Investigadoras''. M. G. Villanueva-Utrilla and Humberto Salazar thank the Vicerrectoría de Investigación y Estudios de Posgrado for support through the ``Centro Interdisciplinario de Investigación y Enseñanza de la Ciencia'' (CIIEC). Marcela Marín and R. Gaitán also acknowledge support from the PAPIIT project IN105825.


\bibliographystyle{JHEP-mod.bst}
\bibliography{references.bib}

\end{document}